\documentclass[amsmath,aps,showpacs,a4paper,10pt]{revtex4}

 \usepackage{epsf}
 \usepackage{graphicx}    

\begin{document}

 \newcommand{\be}[1]{\begin{equation}\label{#1}}
 \newcommand{\ee}{\end{equation}}
 \newcommand{\bea}{\begin{eqnarray}}
 \newcommand{\eea}{\end{eqnarray}}
 \def\disp{\displaystyle}

 \def\gsim{ \lower .75ex \hbox{$\sim$} \llap{\raise .27ex \hbox{$>$}} }
 \def\lsim{ \lower .75ex \hbox{$\sim$} \llap{\raise .27ex \hbox{$<$}} }

 \begin{titlepage}

 \begin{flushright}
 arXiv:1002.4230
 \end{flushright}

 \title{\Large \bf Spinor Dark Energy and Cosmological
 Coincidence Problem}

 \author{Hao~Wei\,}
 \email[\,email address:\ ]{haowei@bit.edu.cn}
 \affiliation{Department of Physics, Beijing Institute
 of Technology, Beijing 100081, China}

 \begin{abstract}\vspace{1cm}
 \centerline{\bf ABSTRACT}
 Recently, the so-called Elko spinor field has been proposed
 to be a candidate of dark energy. It is a non-standard
 spinor and has unusual properties. When the Elko spinor field
 is used in cosmology, its unusual properties could bring some
 interesting consequences. In the present work, we discuss
 the cosmological coincidence problem in the spinor dark
 energy models by using the dynamical system method. Our
 results show that the cosmological coincidence problem should
 be taken to heart in the investigations of spinor dark energy
 models.
 \end{abstract}


 \maketitle

 \end{titlepage}

 \renewcommand{\baselinestretch}{1.1}


\section{Introduction}\label{sec1}

Since the discovery of accelerated expansion of our universe,
 dark energy has been one of the most active fields in modern
 cosmology~\cite{r1,r2,r3}. The simplest candidate of dark
 energy is a tiny positive cosmological constant. As an
 alternative to the cosmological constant, some dynamical
 field models have been proposed. These dynamical field models
 can be categorized into three major types: (complex) scalar
 field models (e.g. quintessence~\cite{r4,r5}, phantom~\cite{r6},
 k-essence~\cite{r7}, quintom~\cite{r8,r9,r10},
 hessence~\cite{r11,r12}), vector field models
 (e.g.~\cite{r13,r14,r15,r16}), and spinor field models. Of
 course, there are also other dark energy models which are not
 directly described by quantum fields, and so we do not
 mention them here.

To our knowledge, in the literature there are relatively few
 works on the dark energy models with spinor fields. In~\cite{r17},
 the Bianchi type~I cosmology with Dirac spinor fields has been
 investigated. In~\cite{r18}, it is found that the Dirac spinor
 fields could be responsible for the cosmic acceleration.
 In~\cite{r19}, the massive non-linear dark spinors have been
 discussed. In~\cite{r20}, the spinor quintom has been studied.

It is worth noting that all the spinors considered in the
 aforementioned works~\cite{r17,r18,r19,r20} are Dirac spinors.
 In fact, there is a different type of spinor in the literature,
 namely, the so-called Elko spinor (e.g.~\cite{r21,r29}), which
 is similar to Majorana spinor. In the beginning, the Elko
 spinor was considered as a candidate of dark matter~\cite{r21}.
 Subsequently, it has been used to drive
 inflation~\cite{r22,r23,r24,r25,r26}. Recently, the Elko spinor
 has been proposed to be a candidate of dark energy~\cite{r27}.
 In fact, this type of dark energy model described by the Elko
 spinor fields is the one we will discuss in the present work.

Following~\cite{r22,r23,r24,r27}, here is a brief review of the
 so-called Elko spinor. It is a spin one half field with mass
 dimension one~\cite{r21}. Unlike the standard fields which
 obey $(CPT)^2=1$, the Elko spinor is non-standard spinor
 according to the Wigner classification~\cite{r28} and obeys
 the unusual property $(CPT)^2=-1$ instead. In fact, the Elko
 spinor fields (together with Majorana spinor fields) belong
 to a wider class of spinor fields, i.e., the so-called
 flagpole spinor fields, according to the Lounesto general
 classification of all spinor fields~\cite{r29,r36}. The Elko
 spinors are defined by~\cite{r21,r24,r25}
 \be{eq1}
 \lambda=\left(
  \begin{array}{c}
  \pm\sigma_2\phi^\ast_L \\
  \phi_L\end{array}
 \right),
 \ee
 where subscript $L$ refers to left-handed spinor; $\sigma_2$
 denotes the second Pauli matrix; $\phi^\ast_L$ denotes the
 complex conjugate of $\phi_L$. Note that the helicities of
 $\phi_L$ and $\sigma_2\phi^\ast_L$ are opposite~\cite{r21}.
 Therefore, there are two distinct helicity configurations
 denoted by $\lambda_{\{-,+\}}$ and $\lambda_{\{+,-\}}$. The
 corresponding action is given by~\cite{r24,r25}
 \be{eq2}
 S=\frac{1}{2}\int\left[\,g^{\mu\nu}{\cal D}_{(\mu}
 \stackrel{\neg}{\lambda}{\cal D}_{\nu)}\lambda
 -V\left(\stackrel{\neg}{\lambda}\!\lambda\right)\right]
 \sqrt{-g}\,\,d^4 x\,,
 \ee
 where $V$ is the potential; the round subscript brackets
 denote symmetrization; ${\cal D}_\mu$ is covariant derivative
 and $\stackrel{\neg}{\lambda}$ is the Elko dual which is
 different from the standard model spinors (see
 e.g.~\cite{r21,r25} for definitions). We consider a spatially
 flat Friedmann-Robertson-Walker (FRW) universe and assume that
 the spinor fields are homogeneous. Following~\cite{r22,r23,r24},
 one can find that
 \be{eq3}
 \lambda_{\{-,+\}}=\phi(t)\frac{\xi}{\sqrt{2}}\,,~~~~~~~
 \lambda_{\{+,-\}}=\phi(t)\frac{\zeta}{\sqrt{2}}\,,
 \ee
 where $\phi$ is a homogeneous real scalar; $\xi$ and $\zeta$
 are constant spinors satisfying
 $\stackrel{\neg}{\xi}\!\xi=\,\stackrel{\neg}{\zeta}\!\zeta=+2$.
 In~\cite{r22,r24,r27}, the effective pressure and energy
 density of the Elko spinor field are found to be
 \bea
 p_\phi=\frac{1}{2}\dot{\phi}^2-V(\phi)
 +\frac{1}{8}H^2\phi^2,\label{eq4}\\
 \rho_\phi=\frac{1}{2}\dot{\phi}^2+V(\phi)
 -\frac{3}{8}H^2\phi^2,\label{eq5}
 \eea
 where $H\equiv\dot{a}/a$ is the Hubble parameter;
 $a=(1+z)^{-1}$ is the scale factor (we have set $a_0=1$); $z$
 is the redshift; a dot denotes the derivatives with respect
 to cosmic time $t$; the subscript ``0'' indicates the present
 value of the corresponding quantity; we use the units
 $\hbar=c=1$. Recently, in~\cite{r27} the Elko spinor field has
 been proposed to be a candidate of dark energy, and we will
 call it ``spinor dark energy'' in the present work. However,
 very recently, it is found that the previous researches
 (e.g.~\cite{r22,r24,r27}) overlooked one part of
 the energy-momentum tensor which arises when the spin
 connection is varied appropriately with respect to the
 metric~\cite{r37,r38}. Therefore, the correct pressure and
 energy density of spinor dark energy should be~\cite{r38}
 \bea
 && p_\phi=\frac{1}{2}\dot{\phi}^2-V(\phi)
 -\frac{3}{8}H^2\phi^2-\frac{1}{4}\dot{H}\phi^2-
 \frac{1}{2}H\phi\dot{\phi}\,,\label{eq6}\\
 && \rho_\phi=\frac{1}{2}\dot{\phi}^2+V(\phi)
 +\frac{3}{8}H^2\phi^2.\label{eq7}
 \eea
 Correspondingly, the equation-of-state parameter
 (EoS) of spinor dark energy reads
 \be{eq8}
 w_\phi\equiv\frac{p_\phi}{\rho_\phi}=
 \frac{\frac{1}{2}\dot{\phi}^2-V(\phi)-\frac{3}{8}H^2\phi^2
 -\frac{1}{4}\dot{H}\phi^2-\frac{1}{2}H\phi\dot{\phi}}
 {\frac{1}{2}\dot{\phi}^2+V(\phi)+\frac{3}{8}H^2\phi^2}\,.
 \ee
 In this case, it is easy to see that
 $w_\phi\ge -1$ when
 $\dot{\phi}^2\ge(\dot{H}\phi^2)/4+H\phi\dot{\phi}/2$, whereas
 $w_\phi<-1$ when
 $\dot{\phi}^2<(\dot{H}\phi^2)/4+H\phi\dot{\phi}/2$. The EoS of
 spinor dark energy crosses the phantom divide $w_{de}=-1$ when
 $\dot{\phi}^2=(\dot{H}\phi^2)/4+H\phi\dot{\phi}/2$.

This note is organized as follows. In Sec.~\ref{sec2},
 we discuss the cosmological coincidence problem in the
 spinor dark energy models by using the dynamical system
 method. A brief summary is given in Sec.~\ref{sec3}. It
 is worth noting that in the present work, we
 merely consider the spinor dark energy with the correct
 pressure and energy density given in Eqs.~(\ref{eq6})
 and~(\ref{eq7}) \cite{r39}.


\section{Spinor dark energy and cosmological coincidence
 problem}\label{sec2}

The cosmological coincidence problem~\cite{r1,r2,r3} is asking
 why are we living in an epoch in which the dark energy
 density and the matter energy density are comparable? Since
 their densities scale differently with the expansion of the
 universe, there should be some fine-tunings. Most dark
 energy models are plagued with this coincidence problem.
 However, this problem can be alleviated in these models via
 the method of scaling solution. If there is a possible
 interaction between dark energy and matter, their evolution
 equations could be rewritten as a dynamical system~\cite{r30}
 (see also e.g.~\cite{r5,r9,r12,r14,r31,r32,r33,r34,r35}).
 There might be some scaling attractors in this dynamical
 system, and both the densities of dark energy and matter are
 non-vanishing constants over there. The universe will
 eventually enter these scaling attractors regardless of the
 initial conditions, and hence the coincidence problem could be
 alleviated without fine-tuning. This method works fairly well
 in most of dark energy models (especially in the scalar field
 models). To our knowledge, there is no attempt to do this in
 spinor dark energy model. Let us have a try.


\subsection{Dynamical system}\label{sec2a}

We consider a flat FRW universe containing both spinor dark
 energy and background matter. The background matter is
 described by a perfect fluid with barotropic EoS, namely
 \be{eq9}
 p_m=w_m\rho_m\equiv (\gamma-1)\rho_m\,,
 \ee
 where the so-called barotropic index $\gamma$ is a positive
 constant. In particular, $\gamma=1$ and $4/3$ correspond to
 dust matter and radiation, respectively. Of course, the
 Friedmann equation and Raychaudhuri equation are given by
 \bea
 &&H^2=\frac{\kappa^2}{3}\rho_{tot}=
 \frac{\kappa^2}{3}\left(\rho_\phi+\rho_m\right),\label{eq10}\\
 &&\dot{H}=-\frac{\kappa^2}{2}\left(\rho_{tot}+p_{tot}\right)
 =-\frac{\kappa^2}{2}\left(\rho_\phi+
 \rho_m+p_\phi+p_m\right),\label{eq11}
 \eea
 where $\kappa^2\equiv 8\pi G=M_{pl}^{-2}$ and $M_{pl}$ is the
 reduced Planck mass. We assume that spinor dark energy and
 background matter interact through a coupling term $Q$,
 according to
 \bea
 &&\dot{\rho}_\phi+3H\left(\rho_\phi
 +p_\phi\right)=-Q\,,\label{eq12}\\
 &&\dot{\rho}_m+3H\left(\rho_m+p_m\right)=Q\,,\label{eq13}
 \eea
 which preserves the total energy conservation equation
 $\dot{\rho}_{tot}+3H\left(\rho_{tot}+p_{tot}\right)=0$.
 Obviously, $Q=0$ corresponds to {\em no} interaction between
 spinor dark energy and background matter.

Following e.g.~\cite{r5,r9,r12,r14,r31,r32,r33,r34,r35}, we
 introduce following dimensionless variables
 \be{eq14}
 x\equiv\frac{\kappa\dot{\phi}}{\sqrt{6}H}\,,~~~~~~~
 y\equiv\frac{\kappa\sqrt{V}}{\sqrt{3}H}\,,~~~~~~~
 u\equiv\frac{\kappa\phi}{2\sqrt{2}}\,,~~~~~~~
 v\equiv\frac{\kappa\sqrt{\rho_m}}{\sqrt{3}H}\,.
 \ee
 Then, we can recast the Friedmann equation~(\ref{eq10}) as
 \be{eq15}
 x^2+y^2+u^2+v^2=1\,.
 \ee
 From Eqs.~(\ref{eq10}), (\ref{eq11}) and (\ref{eq6}), (\ref{eq7}),
 it is easy to find that
 \be{eq16}
 s\equiv -\frac{\dot{H}}{H^2}=3x^2+su^2-\sqrt{3}\,xu+
 \frac{3}{2}\gamma v^2,
 \ee
 in which $s$ appears in both sides. One can solve Eq.~(\ref{eq16})
 and get
 \be{eq17}
 s=\left(3x^2-\sqrt{3}\,xu+\frac{3}{2}\gamma v^2\right)
 \left(1-u^2\right)^{-1}.
 \ee
 By the help of Eqs.~(\ref{eq10}), (\ref{eq11}) and (\ref{eq6}),
 (\ref{eq7}), the evolution equations (\ref{eq12}) and
 (\ref{eq13}) can be rewritten as a dynamical system, namely
 \bea
 &&x^\prime=(s-3)x+\frac{\sqrt{3}}{2}u
 -\frac{\kappa V_{,\phi}}{\sqrt{6}H^2}-Q_1\,,\label{eq18}\\
 &&y^\prime=sy+\frac{x}{\sqrt{2}H}
 \frac{V_{,\phi}}{\sqrt{V}}\,,\label{eq19}\\
 &&u^\prime=\frac{\sqrt{3}}{2}x\,,\label{eq20}\\
 &&v^\prime=\left(s-\frac{3}{2}\gamma\right)v+Q_2\,,\label{eq21}
 \eea
 where
 \be{eq22}
 Q_1\equiv\frac{\kappa Q}{\sqrt{6}H^2\dot{\phi}}\,,~~~~~~~
 Q_2\equiv\frac{vQ}{\,2H\rho_m}\,,
 \ee
 a prime and the subscript ``$,\phi$'' denote derivatives with
 respect to $N\equiv\ln a$ and $\phi$, respectively; we have
 used the universal relation $f^\prime=H^{-1}\dot{f}$ for any
 function $f$. On the other hand, the fractional energy densities
 $\Omega_i\equiv (\kappa^2\rho_i)/(3H^2)$ of spinor dark energy
 and background matter are given by
 \be{eq23}
 \Omega_\phi=x^2+y^2+u^2,~~~~~~~
 \Omega_m=v^2.
 \ee
 The EoS of spinor dark energy reads
 \be{eq24}
 w_\phi=\frac{p_\phi}{\rho_\phi}=
 \frac{x^2-y^2-u^2+\frac{2}{3}su^2-\frac{2}{\sqrt{3}}\,xu}
 {x^2+y^2+u^2}\,.
 \ee

Eqs.~(\ref{eq18})---(\ref{eq21}) could be an autonomous
 system when the potential $V(\phi)$ and the interaction
 term $Q$ are chosen to be suitable forms. In fact, we will
 consider the model with a power-law or exponential potential
 in the next subsections. In each model with different
 potential, we consider four cases with various interaction
 forms between spinor dark energy and background matter. The
 first case is the one without interaction, i.e., $Q=0$. The
 other three cases are taken as the most familiar interaction
 terms extensively considered in the literature, namely
 \begin{eqnarray*}
 &{\rm Case~(I)} &Q=0\,,\\
 &{\rm Case~(II)} &Q=\alpha\kappa\rho_m\dot{\phi}\,,\\
 &{\rm Case~(III)} &Q=3\beta H\rho_{tot}
 =3\beta H\left(\rho_\phi+\rho_m\right),\\
 &{\rm Case~(IV)} &Q=3\eta H\rho_m\,,
 \end{eqnarray*}
 where $\alpha$, $\beta$ and $\eta$ are dimensionless
 constants. The interaction form Case~(II) arises from, for
 instance, string theory or scalar-tensor theory (including
 Brans-Dicke theory)~\cite{r31,r32,r33}. The interaction
 forms Case~(III)~\cite{r34} and Case~(IV)~\cite{r35} are
 phenomenally proposed to alleviate the coincidence problem
 in the other dark energy models.


\subsection{Spinor dark energy with a power-law
 potential}\label{sec2b}

In this subsection, we consider the spinor dark energy model
 with a power-law potential
 \be{eq25}
 V(\phi)=V_0\left(\kappa\phi\right)^n,
 \ee
 where $n$ is a dimensionless constant. Actually, in most of
 the models with Elko spinor
 field~\cite{r21,r22,r23,r24,r25,r26,r27}, the potential is
 usually chosen to be, for instance, $\frac{1}{2}m^2\phi^2$
 or $\alpha\phi^4$. It is easy to see that they are the
 special cases of the power-law potential considered here. In
 this case, Eqs.~(\ref{eq18})---(\ref{eq21}) become
 \bea
 &&x^\prime=(s-3)x+\frac{\sqrt{3}}{2}u
 -\frac{\sqrt{3}}{4}ny^2u^{-1}-Q_1\,,\label{eq26}\\
 &&y^\prime=sy+\frac{\sqrt{3}}{4}nxyu^{-1},\label{eq27}\\
 &&u^\prime=\frac{\sqrt{3}}{2}x\,,\label{eq28}\\
 &&v^\prime=\left(s-\frac{3}{2}\gamma\right)v+Q_2\,.\label{eq29}
 \eea
 If $Q$ is given, we can obtain the critical
 points $(\bar{x},\bar{y},\bar{u},\bar{v})$ of
 the autonomous system by imposing the conditions
 $\bar{x}^\prime=\bar{y}^\prime=\bar{u}^\prime=\bar{v}^\prime=0$.
 Of course, they are subject to the Friedmann constraint
 Eq.~(\ref{eq15}), i.e.,
 $\bar{x}^2+\bar{y}^2+\bar{u}^2+\bar{v}^2=1$. One the other
 hand, by definitions in Eq.~(\ref{eq14}), $\bar{x}$,
 $\bar{y}$, $\bar{u}$,  $\bar{v}$ should be real, and
 $\bar{y}\ge 0$, $\bar{v}\ge 0$ are required.

Here we consider the interaction forms $Q$ given in the end of
 Sec.~\ref{sec2a} one by one. In Case~(I) $Q=0$, the corresponding
 $Q_1=Q_2=0$. There is only one critical point
 $\{\bar{x}=0,\ \bar{y}=\sqrt{\frac{2}{2+n}},\ \bar{u}=
 \pm\sqrt{\frac{n}{2+n}},\ \bar{v}=0\}$. This is not a scaling
 solution because its $\Omega_m=\bar{v}^2=0$. Therefore, the
 coincidence problem persists. In Case~(II)
 $Q=\alpha\kappa\rho_m\dot{\phi}$, the corresponding
 $Q_1=\sqrt{\frac{3}{2}}\alpha v^2$, and
 $Q_2=\sqrt{\frac{3}{2}}\alpha xv$. There are two critical points.
 The first one is given by
 \be{eq30}
 \left\{\bar{x}=0,\ \bar{y}=0,
 \ \bar{u}=\frac{-1+\sqrt{1+8\alpha^2}}{2\sqrt{2}\,\alpha},
 \ \bar{v}=\frac{1}{2}\sqrt{\frac{-1+\sqrt{1+8\alpha^2}}
 {\alpha^2}}\,\right\},
 \ee
 which is a scaling solution. The second one is
 $\{\bar{x}=0,\ \bar{y}=\sqrt{\frac{2}{2+n}},\ \bar{u}=
 \pm\sqrt{\frac{n}{2+n}},\ \bar{v}=0\}$, which is not a scaling
 solution. Finally, we find that in both Cases~(III)
 $Q=3\beta H\rho_{tot}$ and (IV) $Q=3\eta H\rho_m$,
 there is {\em no} critical point and hence there is {\em no}
 attractor of course. Therefore, the cosmological evolution
 trajectory completely depends on the initial conditions, and
 the coincidence problem is inevitable. The fine-tuning of
 initial conditions is required.

So, the only hope to alleviate the coincidence problem relies
 on the sole scaling solution given in Eq.~(\ref{eq30}) for
 Case~(II) $Q=\alpha\kappa\rho_m\dot{\phi}$. However, its
 stability is required in order to be an attractor which is
 necessary to alleviate the coincidence problem. To study the
 stability of the critical points for the autonomous system
 Eqs.~(\ref{eq26})---(\ref{eq29}), we substitute linear
 perturbations $x\to\bar{x}+\delta x$,
 $y\to\bar{y}+\delta y$, $u\to\bar{u}+\delta u$, and
 $v\to\bar{v}+\delta v$ about the critical point
 $(\bar{x},\bar{y},\bar{u},\bar{v})$ into the autonomous system
 Eqs.~(\ref{eq26})---(\ref{eq29}) and linearize them. Because
 of the Friedmann constraint~(\ref{eq15}), there are only three
 independent evolution equations, namely
 \bea
 &&\delta x^\prime=(\bar{s}-3)\delta x+\bar{x}\delta s+
 \frac{\sqrt{3}}{2}\,\delta u-
 \frac{\sqrt{3}}{4}n\left(2\bar{y}\bar{u}^{-1}\delta y-
 \bar{y}^2 \bar{u}^{-2}\delta u\right)-\delta Q_1\,,\label{eq31}\\
 &&\delta y^\prime=\bar{s}\delta y+\bar{y}\delta s
 +\frac{\sqrt{3}}{4}n\left[\bar{u}^{-1}\left(\bar{x}\delta y
 +\bar{y}\delta x\right)-\bar{x}\bar{y}\bar{u}^{-2}\delta u
 \right],\label{eq32}\\
 &&\delta u^\prime=\frac{\sqrt{3}}{2}\,\delta x\,,\label{eq33}
 \eea
 where
 \bea
 &&\bar{s}=\left[3\bar{x}^2-\sqrt{3}\,\bar{x}\bar{u}+
 \frac{3}{2}\gamma\left(1-\bar{x}^2-\bar{y}^2-\bar{u}^2\right)
 \right]\left(1-\bar{u}^2\right)^{-1},\label{eq34}\\
 &&\delta s=\left[2\bar{u}\bar{s}\delta u+6\bar{x}\delta x-
 \sqrt{3}\left(\bar{x}\delta u+\bar{u}\delta x\right)-
 3\gamma\left(\bar{x}\delta x+\bar{y}\delta y+\bar{u}\delta u
 \right)\right]\left(1-\bar{u}^2\right)^{-1},\label{eq35}
 \eea
 and $\delta Q_1$ is the linear perturbation coming from $Q_1$.
 The three eigenvalues of the coefficient matrix of
 Eqs.~(\ref{eq31})---(\ref{eq33}) determine the stability of
 the critical point. For Case~(II)
 $Q=\alpha\kappa\rho_m\dot{\phi}$, the corresponding
 $\delta Q_1=-\sqrt{6}\,\alpha\left(\bar{x}\delta x
 +\bar{y}\delta y+\bar{u}\delta u\right)$. We find that the
 corresponding eigenvalues for the critical point~(\ref{eq30})
 are given by
 \be{eq36}
 \hspace{-2.0mm} 
 \left\{\frac{3\gamma}{2},\,\frac{1}{4}\left[3(\gamma-2)-
 \sqrt{12\sqrt{1+8\alpha^2}+\left[3(\gamma-2)\right]^2}\,\right],
 \,\frac{1}{4}\left[3(\gamma-2)+
 \sqrt{12\sqrt{1+8\alpha^2}+\left[3(\gamma-2)\right]^2}\,\right]
 \right\}.
 \ee
 Note that $3\gamma/2$ is positive, and the second and third
 eigenvalues are negative and positive, respectively. So, the
 critical point given in Eq.~(\ref{eq30}) is unstable, and
 hence it is {\em not} an attractor. Therefore, the hope to
 alleviate the coincidence problem is shattered. Due to the
 failure in the models with a power-law potential, we should
 turn to the models with another potential.


\subsection{Spinor dark energy with an exponential
 potential}\label{sec2c}

In this subsection, we consider the spinor dark energy models
 with an exponential potential
 \be{eq37}
 V(\phi)=V_0\, e^{-\epsilon\kappa\phi},
 \ee
 where $\epsilon$ is a dimensionless constant. In this
 case, Eqs.~(\ref{eq18})---(\ref{eq21}) become
 \bea
 &&x^\prime=(s-3)x+\frac{\sqrt{3}}{2}u
 +\sqrt{\frac{3}{2}}\,\epsilon y^2-Q_1\,,\label{eq38}\\
 &&y^\prime=sy-\sqrt{\frac{3}{2}}\,\epsilon xy\,,\label{eq39}\\
 &&u^\prime=\frac{\sqrt{3}}{2}x\,,\label{eq40}\\
 &&v^\prime=\left(s-\frac{3}{2}\gamma\right)v+Q_2\,.\label{eq41}
 \eea
 If $Q$ is given, we can obtain the critical
 points $(\bar{x},\bar{y},\bar{u},\bar{v})$ of
 the autonomous system by imposing the conditions
 $\bar{x}^\prime=\bar{y}^\prime=\bar{u}^\prime=\bar{v}^\prime=0$.
 Of course, they are subject to the Friedmann constraint
 Eq.~(\ref{eq15}), i.e.,
 $\bar{x}^2+\bar{y}^2+\bar{u}^2+\bar{v}^2=1$. One the other
 hand, by definitions in Eq.~(\ref{eq14}), $\bar{x}$,
 $\bar{y}$, $\bar{u}$,  $\bar{v}$ should be real, and
 $\bar{y}\ge 0$, $\bar{v}\ge 0$ are required.

We consider the four cases of the interaction term $Q$ given
 in the end of Sec.~\ref{sec2a}. Unfortunately, similar to
 the models with a power-law potential, we find that in both
 Cases~(III) and~(IV) there is {\em no} critical point and
 hence there is {\em no} attractor of course. So, the
 cosmological evolution trajectory completely depends on the
 initial conditions, and the coincidence problem is inevitable.
 The fine-tuning of initial conditions is required. In
 Case~(I), there are two critical points, i.e.,
 $\{\bar{x}=0,\ \bar{y}=0,\ \bar{u}=0,\ \bar{v}=1\}$~and
 \be{eq42}
 \left\{\bar{x}=0,\ \bar{y}=\frac{1}{2}
 \sqrt{\frac{-1+\sqrt{1+8\epsilon^2}}{\epsilon^2}},
 \ \bar{u}=\frac{1-\sqrt{1+8\epsilon^2}}{2\sqrt{2}\,\epsilon},
 \ \bar{v}=0\right\}.
 \ee
 Unfortunately, these two critical points are not scaling
 solutions. In Case~(II), there are also two critical points.
 The first one is the same given in Eq.~(\ref{eq30}), which is
 a scaling solution. The second one is the same given in
 Eq.~(\ref{eq42}), which is not a scaling solution.

Again, the only hope to alleviate the coincidence problem relies
 on the sole scaling solution given in Eq.~(\ref{eq30}) for
 Case~(II) $Q=\alpha\kappa\rho_m\dot{\phi}$. To study the
 stability of the critical points for the autonomous system
 Eqs.~(\ref{eq38})---(\ref{eq41}), we substitute linear
 perturbations $x\to\bar{x}+\delta x$,
 $y\to\bar{y}+\delta y$, $u\to\bar{u}+\delta u$, and
 $v\to\bar{v}+\delta v$ about the critical point
 $(\bar{x},\bar{y},\bar{u},\bar{v})$ into the autonomous system
 Eqs.~(\ref{eq38})---(\ref{eq41}) and linearize them. Because
 of the Friedmann constraint~(\ref{eq15}), there are only three
 independent evolution equations, namely
 \bea
 &&\delta x^\prime=(\bar{s}-3)\delta x+\bar{x}\delta s+
 \frac{\sqrt{3}}{2}\,\delta u+\sqrt{6}\,\epsilon\bar{y}\delta y
 -\delta Q_1\,,\label{eq43}\\
 &&\delta y^\prime=\bar{s}\delta y+\bar{y}\delta s
 -\sqrt{\frac{3}{2}}\,\epsilon\left(\bar{x}\delta y+
 \bar{y}\delta x\right),\label{eq44}\\
 &&\delta u^\prime=\frac{\sqrt{3}}{2}\,\delta x\,,\label{eq45}
 \eea
 where $\bar{s}$ and $\delta s$ are given in Eqs.~(\ref{eq34})
 and~(\ref{eq35}), respectively. $\delta Q_1$ is the linear
 perturbation coming from $Q_1$. The three eigenvalues of the
 coefficient matrix of Eqs.~(\ref{eq43})---(\ref{eq45})
 determine the stability of the critical point. For Case~(II)
 $Q=\alpha\kappa\rho_m\dot{\phi}$, the corresponding
 $\delta Q_1=-\sqrt{6}\,\alpha\left(\bar{x}\delta x
 +\bar{y}\delta y+\bar{u}\delta u\right)$. We find that the
 corresponding eigenvalues for the critical point~(\ref{eq30})
 are the same given in Eq.~(\ref{eq36}), in which three
 eigenvalues are positive, negative and positive, respectively.
 So, the critical point given in Eq.~(\ref{eq30}) is also
 unstable for the models with an exponential potential, and
 hence it is {\em not} an attractor. Therefore, the hope to
 alleviate the coincidence problem is shattered again.


\section{Summary}\label{sec3}

Recently, the so-called Elko spinor field has been proposed to
 be a candidate of dark energy. It is a non-standard spinor and
 has unusual properties. When the Elko spinor field is used in
 cosmology, its unusual properties could bring some interesting
 consequences. In this work, we discussed the cosmological
 coincidence problem in the spinor dark energy models by using
 the dynamical system method. According to the results obtained
 in Sec.~\ref{sec2}, we should admit that in the spinor dark
 energy models with $p_\phi$ and $\rho_\phi$ given in
 Eqs.~(\ref{eq6}) and~(\ref{eq7}) coming from~\cite{r38}, it is
 a hard task to alleviate the coincidence problem~\cite{r39}.
 Nevertheless, it is still possible to find some suitable
 potentials $V(\phi)$ and interaction forms $Q$ to obtain the
 scaling attractors of the most general dynamical
 system~\cite{r40}, and hence the hope to alleviate the coincidence
 problem still exists, although this is a fairly hard task.
 Of course, there might be other smart methods different from
 the usual method used in most of dark energy models to
 alleviate the coincidence problem. Anyway, our results
 obtained in the present work showed that the cosmological
 coincidence problem should be taken to heart in the investigations
 of spinor dark energy models.


\section*{ACKNOWLEDGEMENTS}

We thank the anonymous referees for quite useful suggestions,
 which help us to improve this work. We are grateful to
 Professors Rong-Gen~Cai, Shuang~Nan~Zhang, Zong-Hong~Zhu
 and Jian-Min~Wang for helpful discussions. We also thank
 Minzi~Feng, as well as Hongsheng~Zhang, Xiao-Peng~Ma, Bo~Tang,
 for kind help and discussions. This work was supported in
 part by NSFC under Grant No.~10905005, the Excellent Young
 Scholars Research Fund of Beijing Institute of Technology,
 and the Fundamental Research Fund of Beijing Institute of
 Technology.

\renewcommand{\baselinestretch}{1.1}



\begin{thebibliography}{99}
\bibitem{r40}
Here we would like to briefly discuss the most general case.
 From the most general Eq.~(\ref{eq20}), we have $\bar{x}=0$ at
 the critical points. From Eq.~(\ref{eq15}), it is easy to see
 that $1-\bar{u}^2=\bar{y}^2+\bar{v}^2\ge 0$. Therefore, from
 Eq.~(\ref{eq17}), we find that $\bar{s}>0$ when
 $\bar{v}\not=0$ (i.e., $\Omega_m\not=0$, scaling solution),
 whereas $\bar{s}=0$ when $\bar{v}=0$ (i.e., $\Omega_m=0$, dark
 energy dominated). From Eq.~(\ref{eq24}), we see
 that $w_\phi>-1$ when $\bar{v}\not=0$ (and hence $\bar{s}>0$),
 whereas $w_\phi=-1$ when $\bar{v}=0$ (and hence $\bar{s}=0$).
 Therefore, at the critical points (if any), the big rip can
 be avoided because $w_\phi\ge -1$ (and hence $w_{\rm eff}\ge -1$).
\end{thebibliography}
\end{document}